\documentclass[3p,twocolumn]{elsarticle}

\usepackage{amsmath}

\usepackage{hyperref}

\hypersetup{breaklinks=true,
}
\usepackage{graphicx}
\usepackage[dvipsnames]{xcolor}
\usepackage{bbm}
\usepackage{amssymb}

\usepackage[caption=false]{subfig} 

\usepackage{lmodern}
\usepackage[T1]{fontenc}
\usepackage{wasysym}
\usepackage{color}
\usepackage[english]{babel}
\usepackage{gensymb} 
\usepackage{array} 

\graphicspath{{Figs/}}

\usepackage{amssymb}

\newcommand{\correction}{\textcolor{black}}

\begin{document}

\begin{frontmatter}
\title{Ferromagnetic contamination of Ultra-Low-Field-NMR sample containers. Quantification of the problem and possible solutions.
}
\author{Giuseppe Bevilacqua\fnref{DIISM}}
\author{Valerio Biancalana\corref{cor1}\fnref{DIISM}}
\cortext[cor1]{Corresponding author}
\ead{valerio.biancalana@unisi.it}
\author{Marco Consumi\corref{}\fnref{DBCF}}
\author{Yordanka Dancheva\fnref{Aerospazio}}
\author{Claudio Rossi\fnref{DBCF}}
\author{Leonardo Stiaccini\fnref{DSFTA}}
\author{Antonio Vigilante\fnref{DIISM,UCL}}

\address[DIISM]{Dept. of Information Engineering and Mathematics - DIISM, University of Siena - Via Roma 56 53100 Siena, Italy}
\address[DBCF]{Dept. of Biotechnology, Chemistry and Pharmacy - DBCF, University of Siena - Via A.Moro 2, 53100 Siena, Italy}
\address[Aerospazio]{currently at: Aerospazio Tecnologie srl, Strada di Ficaiole, 53040 Rapolano Terme (SI), Italy}
\address[DSFTA]{Dept. of Physical Sciences, Earth and Environment - DSFTA, University of Siena - Via Roma 56, 53100 Siena,  Italy}
\address[UCL]{Currently at: Department of Physics and Astronomy, University College London, Gower Street, London WC1E 6BT, United Kingdom}

\begin{abstract}
  The presence of a weak remanence in Ultra-Low-Field (ULF) NMR sample containers is investigated on the basis of proton precession. The high-sensi\-ti\-vi\-ty magnetometer used for the NMR detection,  enables simultaneously the measurement of the static field produced in the sample proximity by ferromagnetic contaminants. The presence of the latter is studied by high resolution chemical analyses of the surface, based on X-ray fluorescence spectroscopy and secondary ions mass spectroscopy. Methodologies to reduce the contamination are explored and characterized. This study is of relevance in any ULF-NMR experiment, as in the ULF regime spurious ferromagnetism becomes easily a dominant cause of artefacts.
\end{abstract}


\end{frontmatter}


\section{Introduction}

Most of the magnetic resonance (MR) experiments aimed at spectroscopic measurements or at imaging (MRI) suffer from  distortions of  the magnetic field. Hence, particular care is necessary to prevent field disturbance induced  in the vicinity of the sample by parts of the apparatus, including the sample itself. 

In conventional (high field) NMR experiments, field distortions causing line broadening or misshaping
(in  spectroscopy) as well as image artifacts (in MRI) arise most commonly from susceptibility (see e.g. Refs.~\cite{wapler_jmr_14,doty_cmr_98} and references therein) and conductance \cite{bennett_jap_96,moessle_jmr_06}, but also from ferromagnetic terms \cite{doty_cmr_98}. 

In earth-field \cite{zampetoulas_jmr_17}, ultra-low-field (ULF) \cite{biancalana_arnmrs_13,tayler_rsi_17}, and zero-field \cite{barskiy_nat_19} MR apparatuses,  the ferromagnetic terms may play an important --potentially dominant-- role, and their presence can be detected directly by the same sensor that measures the MR signal \cite{biancalana_rsi_19,tayler_apl_19}. This aspect is the focus of the present work. In particular, we study, characterize and analyze spurious effects occurring in an ULF-NMR apparatus, originating from ferromagnetic contamination of polymeric sample containers (cartridges). These cartridges contain samples for a remote-polarization NMR experiment that uses an optical atomic magnetometer (OAM) as a high-sensitivity, non-inductive detector \cite{biancalana_DH_jpcl_17,biancalana_IDEA_prappl_19,biancalana_apl_19}. 

Magnetic detectors based on OAMs are an interesting class of sensors that rival with the top-sensitivity ones based on superconducting quantum-interference devices (SQUIDs), compared to which they have advantages in terms of maintenance cost, practicality (no cryogenics needed), and of robustness with respect to strong fields.
In some implementations, OAMs have a broadband response, extending to static signals. This feature is here exploited to detect simultaneously both the DC signal due to ferromagnetic impurities and the time-dependent signal generated by nuclear precession.

This work is about the characterization of spurious ferromagnetic remanence of the cartridges and of the material used for their production. Besides magnetometric measurements, surface analysis based on X-ray fluorescence spectroscopy (XRFS) and Time-of-Flight Secondary-Ion Mass-Spectrometry (ToF-SIMS) have been performed. In addition, several  approaches attempted  to remove or reduce the contamination are described and discussed, drawing conclusions about their effectiveness.

Detection and analysis of contamination by ferromagnetic impurities, as well as tiny ferromagnetic behaviour due to specific phenomena, are topics of interest among a wide community, spanning from material science \cite{sepioni_epl_12,wang_jap_14}, to semiconductor technology \cite{kuroda_nat_14}, nanotechnology \cite{zhao_prb_08}, and medicine \cite{muluaka_ijomeh_03}. A recent paper addresses the problem of  characterizing and counteracting ferromagnetic contamination in a variety of metal-oxide substrates, in which weak extrinsic ferromagnetic behaviour is observed \cite{pereira_jpd_11}, also evidencing possible artifacts due to the measurement procedure.

Works devoted to investigate extremely weak ferromagnetic response in nano-structures \cite{zhao_prb_08,wang_jap_14} or diluted dopants \cite{sepioni_epl_12}  commonly make use of  state-of-art magnetometers (typically SQUIDs) and study the saturation level and the hysteresis that characterize the material. On the other hand our measurements are more tightly focused to the implications of spurious ferromagnetism in ULF-NMR apparatuses and are mainly concerned with the remanence. 

The paper is organized as follows: the Sec.~\ref{sec:setup}  provides a brief description of  the OAM used, of the setup making it suited to detect ULF-NMR signals, and of the instrumentation used to chemically analyze the surface contaminants; the Sec.~\ref{sec:measurements} shows the effects caused by the spurious magnetization of the  containers and complementary methodologies used to evaluate and characterize the contamination; the Secs.~\ref{sec:titanium} and \ref{sec:cleaning}  provide the results obtained by the application of methods aimed at preventing/removing  the ferromagnetic contamination, and the information inferred about the bulk or surface localization of the problem. The Sec.~\ref{sec:conclusioni} concludes the paper, providing a summary of the observations and of the results as well as a discussion about possible implications and perspectives.

\section{Materials and methods}
\label{sec:setup}

\subsection{Magnetometer} 
\label{subsec:magnetometer}

\begin{figure}[ht]
   \centering
    \subfloat[\label{fig:setupA}]{
    \includegraphics [angle=0, width=.48\columnwidth]{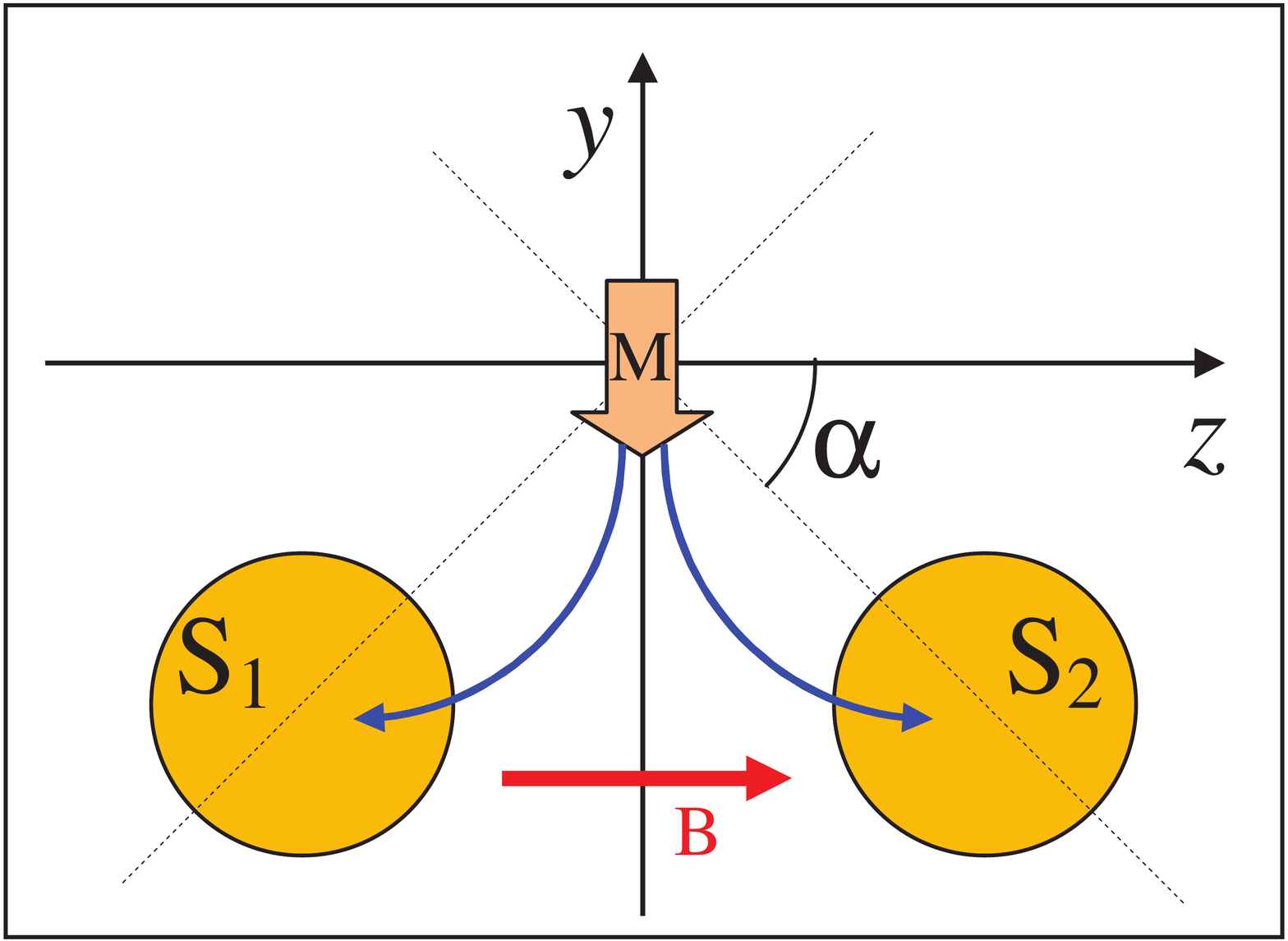}} 
    \hfill
    \subfloat[\label{fig:setupB}]{
    \includegraphics [angle=0, width=.48\columnwidth]{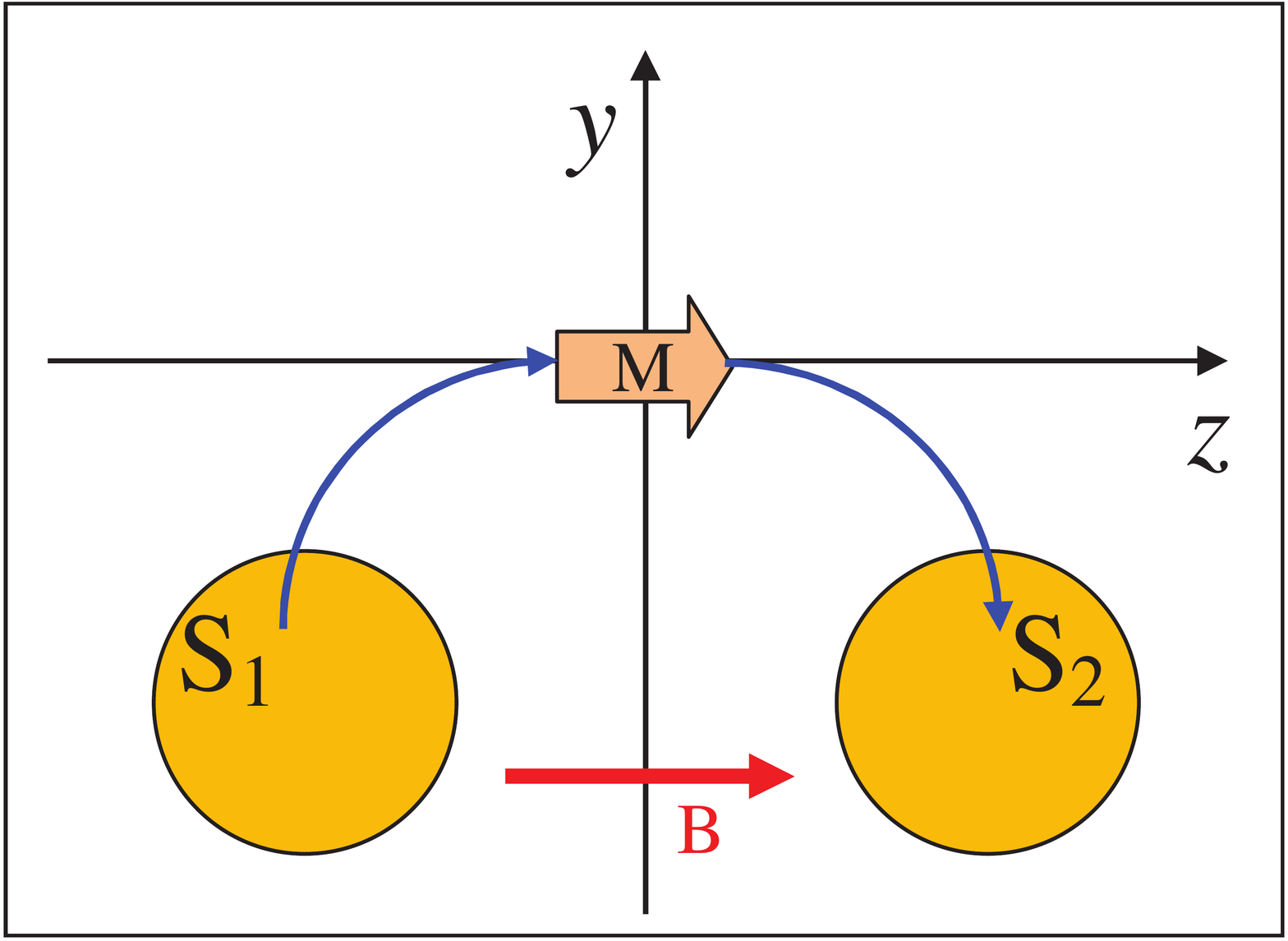}}
     \caption{
	Front view of the detection region. The yellow arrow (M) represents the sample and its magnetization and the circles (S$_1$, S$_2$) represent the two sensors: glass cells containing Cs vapour. Laser beams cross the cell to optically pump the Cs atoms and to probe the atomic state evolution, they  propagate along $x$ and are not represented, $x$ is also the direction of the shuttling system. A static (bias) field (red arrow) is oriented along $z$. Sample and sensors are at different heights, determining the angle $\alpha$ and the consequent sample-sensor coupling factor. The dual magnetometric head   detects  the field generated by the $y$ component of the sample magnetization M as a difference-mode term (a). Such field (blue arrows) in this case is parallel/antiparallel to the static (much stronger) bias field, and contains a time dependent term due to the nuclear precession around B and -possibly- a static term due to the cartridge magnetization. The $x$ and $z$ components of M (b), produce a transverse field perturbation, so to cause only second-order variations of the field modulus. Moreover such variations  are cancelled in the differential measurement, because they appear as a common mode term. 
	}
		\label{fig:setup}
\end{figure}

The magnetometric setup is designed to measure  NMR signals from samples that  have been previously magnetized in a field at the Tesla level, generated by a permanent-magnet Halbach array \cite{halbach_nim_80}. The nuclear precession occurs in a ULF (at micro-Tesla level, corresponding  to proton Larmor frequencies of the order of 100  Hz), so to require non-inductive detection.
The apparatus contains an OAM \cite{biancalana_apb_16} that detects \textit{in situ} the NMR signal, a system for pneumatic sample transfer \cite{biancalana_rsi_14} from the magnetization- to the detection-region, and coils to control \cite{biancalana_rsi_17} and to stabilize \cite{biancalana_rsi_10,biancalana_FPGAstab_prappl_19} the static magnetic field  and to apply the magnetic pulses necessary to manipulate the nuclear spins, making them precessing around the magnetic field direction. 

The whole setup (see Fig.\ref{fig:setup}) is built  around a\correction{n unshielded} dual OAM operating in a Bell \& Bloom configuration \cite{bellandbloom_prl_61} that is described in Ref.~\cite{biancalana_apb_16} and is adapted to ULF-NMR measurement as described in Ref.~\cite{biancalana_zulfJcoupling_jmr_16}. The OAM is a broadband detector, and it may record NMR signals superimposed with- (and modified by-) static terms  generated by the permanent magnetization \cite{biancalana_rsi_19}. The sensor response is nearly flat from DC to about 30Hz. After this cutoff-frequency, the response decreases with a 6dB/oct roll-off \cite{biancalana_apb_16}, while maintaining a nearly constant signal-to-noise ratio for at least 3 octaves. \correction{This feature is due to the magnetic noise floor which constitutes the actual limit to the system sensitivity: the response decrease acts simultaneously on both the signal and on that magnetic disturbances. The intrinsic (non-magnetic) noise terms emerge and start to affect relevantly the S/N ratio above several hundred Hz.}

Briefly, the dual sensor detects the static and time-dependent terms of the magnetic field in two locations nearby the sample position, as sketched in Fig.~\ref{fig:setup}. The scalar nature of the atomic sensors \cite{budker_nat_07}, in the presence of a dominant bias field oriented along a given direction, makes the system responding only to the variations  of the field component along that direction. A small variation of $\delta \vec B$ over a bias field $\vec B_0$ causes a modulus variation of the total field $\delta B_{\mbox{tot}} \approx (\delta \vec B \cdot \vec B_0)/B_0 = \delta B_{\parallel}$~\cite{biancalana_jmr_09}.

As shown in Fig.~\ref{fig:setupA}, in which the bias field is oriented along $z$, a dipolar source --displaced along $y$ over the sensor plane-- produces variations of $B_z$ when its magnetization is oriented along $y$. In this case the two sensors detect opposite $\delta B_{\mbox{tot}}$, so to maximize the dif\-fer\-ence-mode response of the magnetometer output. Disturbances from far located sources  appear as a common-mode term and are  profitably cancelled by difference. The cancellation is improved by an active compensation system. The common mode signal feeds such system \cite{biancalana_FPGAstab_prappl_19}, so to be actively attenuated prior to its cancellation by difference. Attenuation and cancellation occur also for the field induced by sample magnetization along $z$:  as shown in Fig.~\ref{fig:setupB}, the field produced by $M_z$ sums in quadrature to the bias field, so to produce equal variations in the two sensors. In conclusion, only the $y$ component of the sample magnetization produces a detected signal. \correction{It is worth mentioning an important role of the active stabilization system. As the measurement is performed in an unshielded environment, slow drifts of the ambient magnetic field (despite being cancelled by the differerential measurement) would affect the nuclear precession, hindering trace-averaging procedures used to improve the NMR S/N. More specifically, uncompensated field drifts would cause a $T_2$ underestimation in the NMR signal.  }

\subsection{ULF-NMR setup}
\label{subsec:ulf}
In this work we use water samples which, thanks to the proton abundance and long decay time, facilitate the NMR characterization of the sample containers. These containers can be used with other NMR substances, as previously reported \cite{biancalana_DH_jpcl_17,biancalana_zulfJcoupling_jmr_16}. The containers are  sealed cartridges with screwed or glued caps.
The samples are remotely polarized (at about 1 T) and pneumatically shuttled to the detection region, there --after the  application of appropriate spin-tipping field pulses-- the nuclei precess in a  field at microtesla level. To guarantee performance and reliability of the transfer system \cite{biancalana_rsi_14},  an accurate shaping and a sufficient mechanical robustness of the cartridges are required. To this end, an accurate selection of the material is necessary.

Several reasons make the use of metal cartridges disadvantageous.  Despite the low nuclear precession frequencies typical of the ULF regime, eddy currents induced in electrically conductive containers may shield both the  NMR signal  and the spin tipping field. In our setup, the shielding effect hinders particularly the tipping procedure. The latter is performed by means of sudden (non-adiabatic) rotation of the static field, or by the application of resonant pulses. In the resonant case, the species-selectivity is enhanced by applying static field at the hundreds of $\mu$T level together with the ac field pulse, which increases the resonant frequency up to the 10 kHz level: at these frequencies the skin-depth in metals is submillimetric. Moreover, several metallic and metal-alloy materials contain relevant traces of ferromagnetic contaminants  
\cite{lifeng_mpemr_12}, as it was recently pointed out in a ULF-NMR experiment \cite{tayler_apl_19}. Additionally, non-metallic containers may guarantee a wider chemical compatibility with the NMR samples to be analyzed.

\correction{The shuttling system requires precise external sizing and low fragility, making glass a disadvantageous choice.}
Among \correction{other} non-conductive materials, we have selected and charactrized po\-ly\-eth\-er ether ketone (PEEK) for its excellent mechanical properties (machinability, chemical resistance, mechanical strength). 
Some attempts made with other polymeric substances --e.g. Iglidur and Acetale-- led to similar observations. We will briefly deal also with additional tests performed on other materials used to produce polymeric samples with 3D printing.
PEEK finds applications in MRI/NMR setups \cite{wapler_jmr_14}, and --by \correction{v}irtue of its biocompatibility-- is often selected in medicine to replace titanium in implantations, also to avoid artifacts in the MRI post-surgical evaluations~\cite{kurtz_biomat_07}.

When $\delta B_{\mbox{tot}}$ is produced by a NMR sample, the time dependent term is due to nuclear magnetization that precesses in the $xy$ plane, around the $z$ direction, so to have an  $y$ component oscillating at the nuclear Larmor frequency. Permanent magnetization of the sample container appears in the signal as well, with a static difference-mode field variation (DMFV) proportional to the $y$ component of the magnetization.
The $z$ component of the magnetization produces instead first order effects on the precession frequency of nuclei inside the container.

Summarizing, the broadband response of the magnetometer permits to register directly a static signal due to $M_y$, simultaneously with the nuclear precession signal, whose frequency depends on the bias field and is modified by  $M_z$.

Beside this NMR frequency shift, the ferromagnetic contamination causes a broadening of the nuclear resonance \cite{biancalana_rsi_19}. Both NMR shift and broadening can be estimated shot by shot by means of reliable numerical methods \cite{bertocco_ieee_94,yoshida_jpe_82}. 
Repeated (cycled) NMR measurements are commonly performed to improve the signal-to-noise ratio. In our case, as the cartridge undergoes unpredictable rotations during the sample transfer, the effects of the ferromagnetic term appear as distributed widths an\correction{d} shifts of the NMR. These effects can be quantified in terms of: (i) variance of the shot-by-shot frequency estimate, (ii) variance of the shot-by-shot decay-rate estimate, (iii) apparent decay-rate increase in the average trace (where T2* decreases due to fluctuations of the ferromagnetically induced shift, which --as said-- varies with the cartridge rotation angle). More details about these kinds of analyses can be found in ref.\cite{biancalana_rsi_19}

\subsection{Chemical surface analysis}
\label{subsec:surfaceinstrumentation}
Previous measurements \cite{biancalana_rsi_19} performed in our laboratory, differing from other observations \cite{tayler_apl_19}, pointed out that PEEK cartridges show indeed an unexpected magnetization level, which can be evidenced -- with no spatial resolution -- by DC measurements and by ULF NMR spectroscopy. 
A chemical analysis of the sample containers surface provides a complementary insight on the nature of this problem. To this end, ToF-SIMS and XRFS techniques are applied, to extract useful information about the amount and the morphology of the surface contamination. 

XRFS are performed using an Olympus Delta Premium (INNOV-X) instrument to determine major and trace elements, as previously described \cite{leone_cons_19}. The analyses can be performed on the polymeric samples \correction{non-destructively.}

ToF-SIMS measurements are carried out on a TRIFT III spectrometer (Physical Electronics, Chanhassen, MN, USA) equipped with a gold liquid-metal primary ion source by a procedure already reported \cite{leone_ass_12}. There are some geometrical constraints, which limit the size of the analyzed sample. In particular, NMR cartridges do not fit in the holder, and are sacrificed to undergo ToF-SIMS analysis. As an alternative, smaller polymer samples are produced with the same lathing tools.

XRFS and ToF-SIMS analyse small portions of the  polymer surface, so to make their (local) results not necessarily consistent (in terms of estimated amounts) with the (global) magnetometric measurements. The different kinds of analyses provide complementary information and may help in confirming interpretative hypotheses, while performing direct cross-correlation of the observations is not obvious and straightforward, due to the different subjects of measurements (small surface portions, and global surface- or  volume-contaminants, respectively).

Compared to ToF-SIMS, the XRFS has a lower sensitivity (13 ppm, instead of 10ppb-1ppm), but a deeper penetration (several microns instead of few nanometers). ToF-SIMS offers an excellent lateral resolution, which approaches the micrometric level. In addition ToF-SIMS may produce hyperspectral maps, 
where different contaminants can be distinguished with a mass resolution ($m/\Delta m$) of several thousands. Different elements are identified on the basis of the mass to charge ratio ($m/z$).

Chemical images relative to positive ion spectra are acquired with a pulsed, bunched 22 keV Au$^+$ primary ion beam, by rastering the ion beam over a 300~$\mu$m $\times$ 300~$\mu$m  area.  Static ToF-SIMS condition (primary ion dose density $<10^{12}$ ions/cm$^2$) are not requested, so that acquisition time extends to 10 minutes. Positive ion spectra are calibrated with CH$_3^+$ ($m/z=15.023$), C$_2$H$_3^+$ ($m/z=27.023$) C$_3$H$_5^+$ ($m/z=41.039$). Before acquiring spectra, the polymeric samples are maintained overnight in a conditioning pre-chamber, with a vacuum value of $10^{-4}$~Pa. The mass resolution ($m/\Delta m$) is 2000 at $m/z=27$.

\section{Evidence of contamination}

\label{sec:measurements}

Ferromagnetic contaminants on the surface and/or in the polymer bulk, cause a remanence that, after the premagnetization in the Halbach array, makes the cartridge produce
a nearly dipolar field on its exterior and, internally, a field  in average antiparallel to the magnetization and scarcely homogeneous. We are dealing with a spurious remanence that is extremely weak, so to produce only perturbative effects.

Concerning the mentioned dipole approximation, an estimation of the dipolar term and of higher-order multipolar ones, can be derived on the basis of ref.\cite{caciagli_jmmm_18} with the assumption of uniformly distributed contaminants. A multipolar expansion of the field calculated in the proximity of a uniformly, transversely magnetized cylinder of radius R and semi-length L, with magnetization M shows that, indicating with  $z$, $\rho$ and $\varphi$ the radial, axial and azimuthal co-ordinates with respect to the center of the cylinder, the field components are 
\begin{equation}
\begin{split}
 B_z &\approx \frac{1}{2}\frac{\mu_0 M L R^2 \rho z}
{\left( z^2+\rho^2 \right) ^{5/2}}
 \bigl[ 3+\\
&+\frac {5}{8}
\frac {4L^2\left(4z^2-3\rho^2
  \right)-3R^2 \left(4z^2-3\rho^2 \right) }
{ \left( z^2+\rho^2 \right) ^2} \bigr] \cos(\varphi),
\end{split}
\end{equation}
\begin{equation}
\begin{split}
 B_\rho & \approx\frac{1}{2}\frac{\mu_0 M L R^2 }
{\left( z^2+\rho^2 \right) ^{5/2}}
 \Bigl[ \left(2\rho^2 - z^2\right)+ \\
&+\frac {1}{8}
\frac {(3R^2-4L^2) \left(4z^4-27\rho^2z^2+4\rho^4 \right) }
 { \left( z^2+\rho^2 \right) ^2} \Bigr] \cos{\varphi},
\end{split}
\end{equation}
and
\begin{equation}
\begin{split}
 B_\varphi & \approx\frac{1}{2}\frac{\mu_0 M L R^2 }
{\left( z^2+\rho^2 \right) ^{3/2}}
 \Bigl[ 1+ \\
&+\frac {1}{8}
\frac {(4L^2-3R^2) \left(4z^2-\rho^2 \right) }
 { \left( z^2+\rho^2 \right) ^2} \Bigr] \sin(\varphi),
\end{split}
\end{equation}
\\
respectively, where the first lines are dipole terms, qua\-dru\-po\-le terms are missing, and the second lines are octupole terms. On the $z=0$ (equatorial) plane, the axial component vanishes and the two remaining  simplify to
\begin{equation}
 B_\rho \approx \frac{\mu_0 M L R^2 }{\rho^3}
\left[ 1+ \frac{1}{4}\frac{3R^2-4L^2}{\rho^2} \right]\cos(\varphi)
\label{eq:dipolorho}
\end{equation}
\begin{equation}
 B_\varphi \approx \frac{1}{2}\frac{\mu_0 M L R^2 }{\rho^3}
\left[ 1+ \frac{1}{8}\frac{3R^2-4L^2}{\rho^2} \right]\sin(\varphi), 
\label{eq:dipolophi}
\end{equation}
respectively. In the geometry of our setup, where $R/\rho \approx 1/5$ and $L/\rho \approx 1/3$, the dipolar terms exceed the octupolar ones by more than an order of magnitude.

The DMFV depends on the $z$ component of the sample field \correction{at the position of the sensor}, which  (see Fig.\ref{fig:setup}) is given by $B_{\rho}\cos(\alpha)-B_{\varphi}\sin(\alpha)$.
The extreme values of the DMFV observed in large sets of measurements provide a quantitative estimate of the dipolar moment and hence of the average polymer magnetization.

The orientation of the cartridge magnetization is originally parallel to the field generated by the Halbach array, but undergoes an unpredictable, aleatory rotation $\theta$ during the displacement to the measurement region. Thus $\varphi=\theta-\alpha$ is randomly distributed and the field perturbation due to the cartridge magnetization at the measurement stage varies on a shot-by-shot basis.  
As  described in ref.\cite{biancalana_rsi_19}, the response of the magnetometer --which, in turn, is calculated in  \cite{biancalana_apb_16}--  to static field variations can be inferred from the differential phase shift of the polarimetric signals extracted from the two magnetometer outputs.

\begin{figure*}[ht]
   \centering
    \includegraphics [angle=0, width=2\columnwidth]{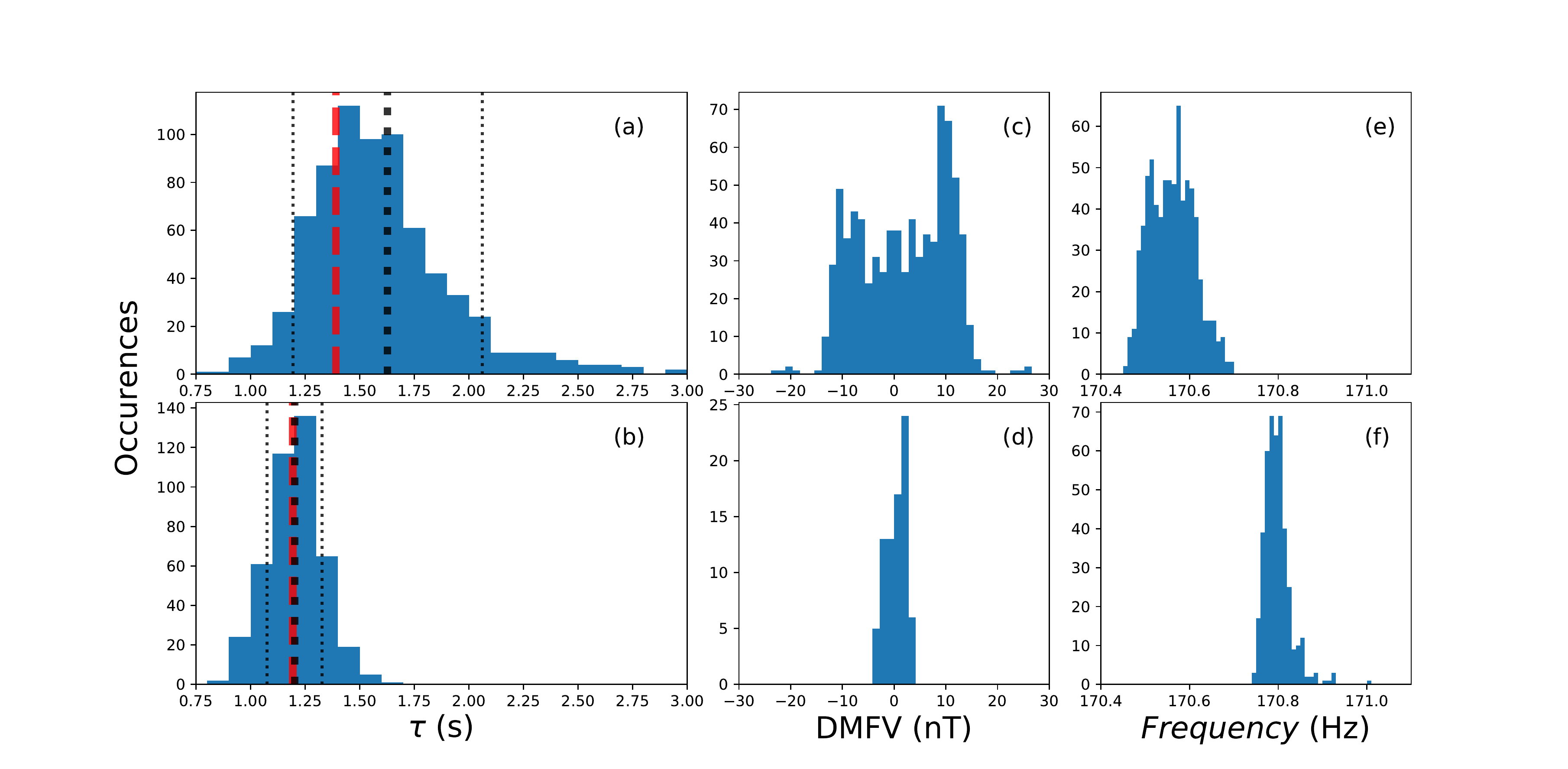}
    \caption{Cartridge characterization in terms of relaxation time (a) and (b),  DMFV distribution (c) and (d), and NMR frequency (e) and (f), over two large sets of measurements. The histogram in the first row (a) (c) and (e) are obtained with a contaminated cartridge, while those in the second row (b) (d) and (f) correspond to a clean cartridge.}
	\label{fig:hists}
\end{figure*}

Fig.\ref{fig:hists} shows histograms of characteristic decay times, DMFVs and proton precession frequencies estimated in two large sets of NMR measurements, prior and after having removed the ferromagnetic contamination (as discussed in the next sections). The first row ((a), (c), (e)) refers to a 804 shots measurement performed on a contaminated cartridge, while the second row ((b), (d), (f)) shows results obtained in 430 shots with a cleaned cartridge (see Sec.\ref{sec:cleaning}). 

\begin{figure*}[ht]
   \centering
    \includegraphics [angle=0, width=2\columnwidth]{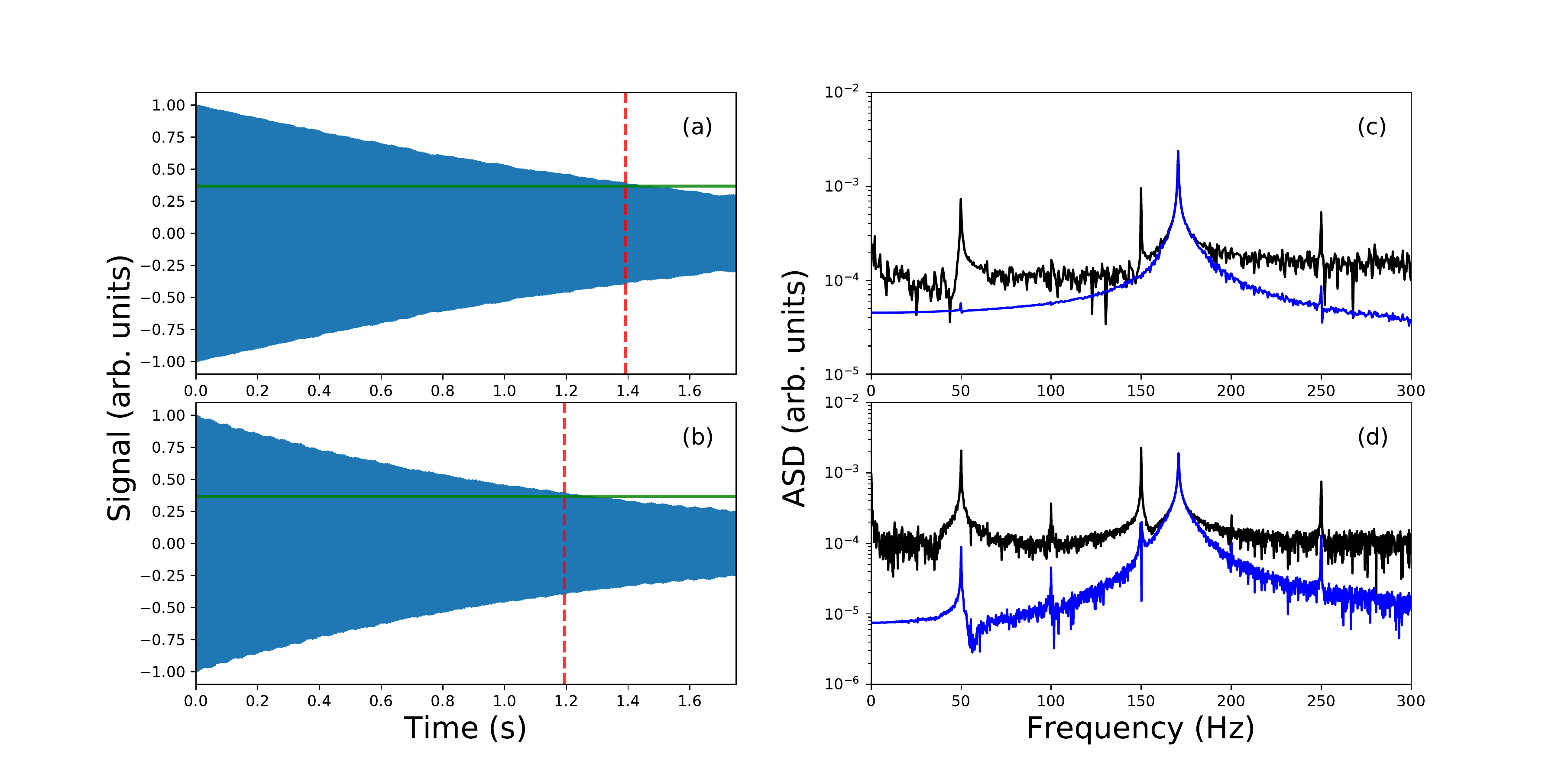}
    \caption{Normalized average traces from 804 NMR measurements with a contaminated cartridge (a), and from 430 measurements with a clean cartridge (b).  The ASD of the corresponding raw data are  shown in the black plots (c) and (d), respectively. A 150 Hz notch filter and a band pass filter around the NMR peak are applied to produce the blue ASD plots and the time-domain traces (a) and (b). In these latter, the green horizontal lines indicate the value $1/e$, and the red-dashed lines indicate the numerically estimates of the decay time: the same as in Fig.\ref{fig:hists} (a) and (b).}
	\label{fig:FIDS}
\end{figure*}

The  characteristic decay time is evaluated both as an average of estimations performed on single traces (thick, black-dot line, with the thinner lines indicating the $\pm$ standarc deviation interval), or directly on the average trace. 
\
The discrepancy between the two estimations is much larger in the case of the contaminated cartridge. This is a direct consequence of the narrower NMR frequency distribution  in the clean sample compared to the contaminated one (cfr. (e) and (f)). A second evidence is that in (b) the  decay rate has narrower distribution compared to (a). It is worth noting that the shortening of the mean decay time observed in  (b) is not relevant. It is due to having reloaded the cartridge  with a different water sample after the cleaning process, not to ferromagnetic contamination. As an example of ULF-NMR measurement results,  Fig.\ref{fig:FIDS} shows averaged traces and corresponding amplitude spectral density (ASD) plots obtained with a contaminated cartridge and a clean one, respectively. These data correspond to the histograms reported in Fig.\ref{fig:hists} (a,c,e) and (b,d,f), respectively. The decay-time  graphical estimates (red-dashed lines) match the numeric evaluations of Fig.\ref{fig:hists} (a) and (b).
\
\
Other tight and quantitative evidences of ferromagnetic contamination emerge from the comparison of measured DMFV distributions (Fig.\ref{fig:hists} (c) and (d)).  The DMFV range decreases in this case from $\pm 15$ to $\pm 3$. Concomitantly,  the distribution of nuclear precession frequencies (Fig.\ref{fig:hists} (e) and (f)), undergoes a range narrowing by a factor of three. 
\
In the case of ordinary cartridges, a DMFV of $\pm40$ nT is typically measured. 
The DMFV is an indicator of magnetic contamination, which, as discussed in the following, occurs --to a large extent-- when the PEEK is lathed. There, uncontrolled and non-reproducible parameters (such as exact tool positioning, cutting angles, tool edge sharpness) introduce an important variability. However, repeated measurements over many sample  permit to estimate typical DMFV values with reasonable accuracy.

The dipole moment, and hence the average magnetization $M$, of the material are evaluated from the DMFV value and from a coupling factor determined by the geometry of the arrangement (eqs.\ref{eq:dipolophi} and \ref{eq:dipolorho}).

In the following we will report the directly measured quantity (i.e. the DMFV), reminding that the dipole is proportional to the DMFV, and hence the magnetization is proportional to the DMFV and inversely proportional to the polymer volume.
In our geometry, where $\rho\approx 50 $ mm, R=9.3 mm, L=16 mm, and $\alpha \approx 45$deg,  the value of M in A/m is about $5\times 10^6$ times the DMFV in nT, i.e. the  $\pm$40 nT DMFV typically measured with cartridges corresponds to an average magnetization of $0.2$~A/m, or (200~$\mu$emu/cm$^3$) (to be doubled, if one considers only the PEEK volume).

Beside the DMFV, the internal field variation inside the cartridges can be estimated from its effects on the NMR signal: as mentioned in Sec.\ref{subsec:magnetometer} and discussed in Ref.~\cite{biancalana_rsi_19}, this feature makes it possible to estimate both the $y$ and the $z$ components of the magnetization  simultaneously. In the following (see Sec. \ref{sec:titanium}) the cartridge magnetization will be linked to a spread of NMR proton frequencies estimated in repeated measurements.
\begin{figure}[t]
   \centering
   \includegraphics [angle=0, width= 0.56 \columnwidth] {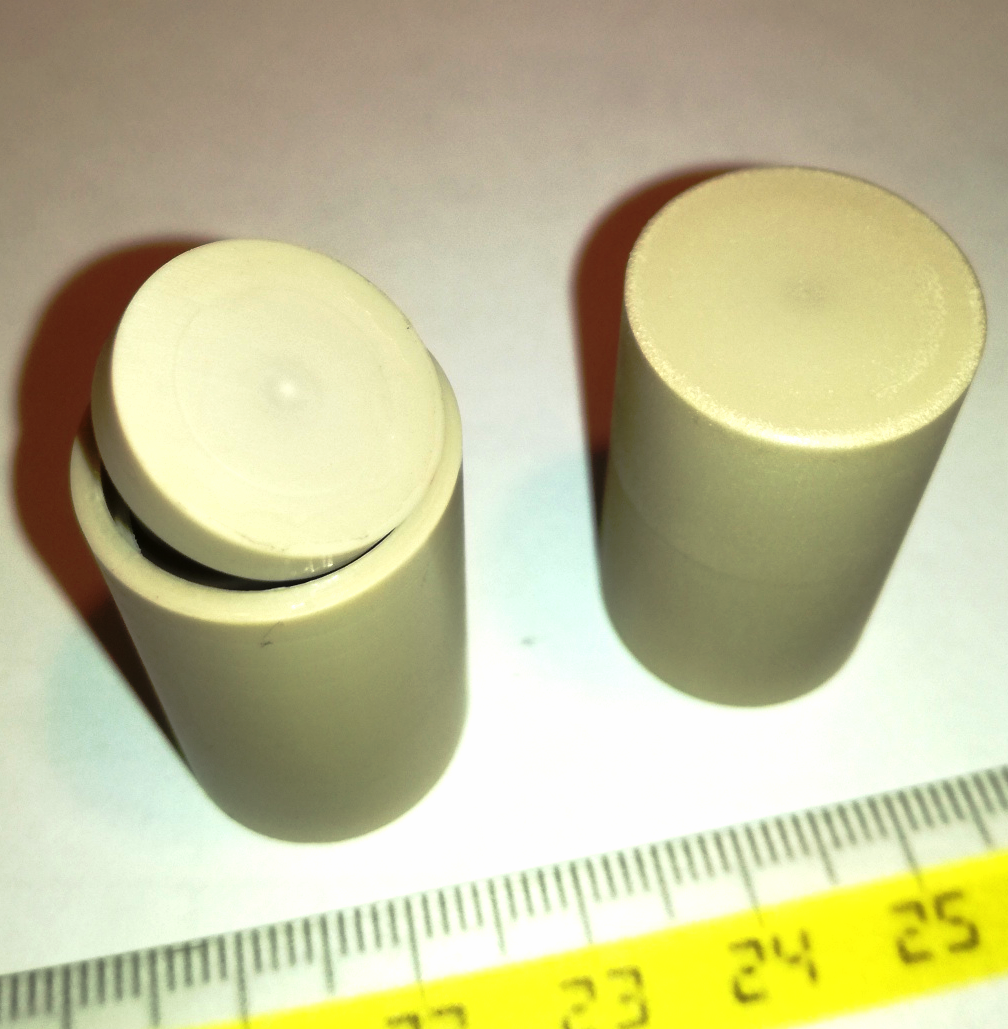}
    \includegraphics [angle=0, width= 0.40 \columnwidth]{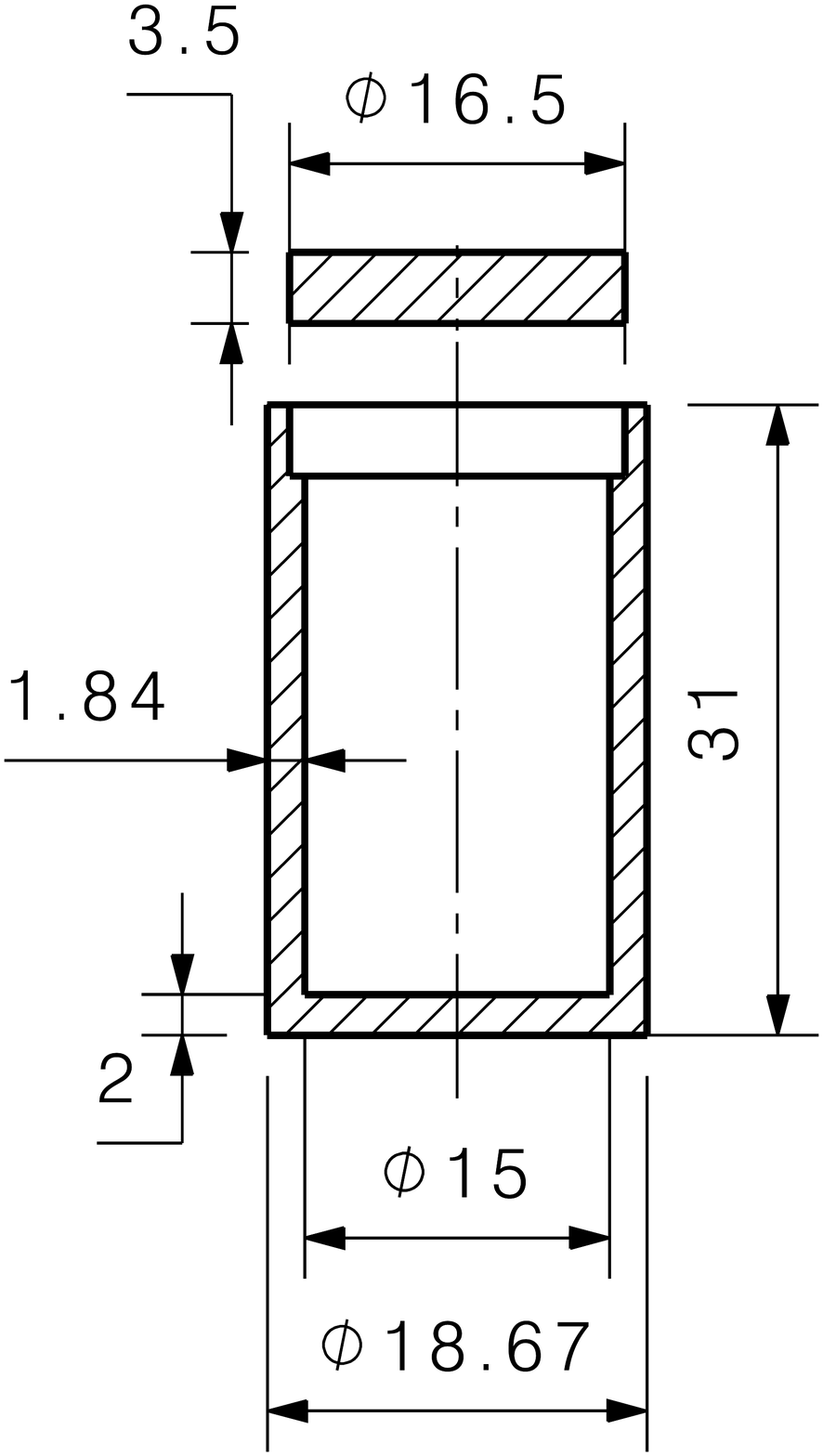}
     \caption{
	The picture shows an empty cartridge and a cylindrical sample, they have identical external size. The relevant dimensions of the cartridge are reported (in mm) in the drawing. The polymer volume of cartridges is 53\% with respect to cylinders, while the machined surface of cylinders is 57\% with respect to cartridges. 
	}
		\label{fig:samples}
\end{figure}

Sets of magnetometric measurements have been performed using cylindrical solid PEEK samples having the same external shape of the cartridges. Hereafter these samples will be referred to as \textit{cylinders}. Both kinds of samples are shown in Fig.~\ref{fig:samples}, together with their relevant sizes. Of course an ULF-NMR characterization is not feasible in the case of cylinders: the magnetometric analysis is limited to the DMFV measurements. 

Typically, cylinders produce half  DMFV than cartridges. 
Being the cylinders about double in polymer volume and half in machined surface compared to cartridges, such DMFV is attributable to a ferromagnetic surface contamination.

This consideration highlights the importance of complementing the magnetometric measurements with spatially and chemically resolved analyses.

XRFS performed on ordinary cartridges and cylinders has provided an immediate evidence of ferromagnetic contaminants, e.g. the first analyzed cylinder (machined with conventional tools) showed the presence of several metallic elements (including ferromagnetic ones), among which:  
Fe (42  ppm), 
Cr (21 ppm),
Cd (18 ppm),
Zn (13 ppm).
XRFS is useful to obtain initial indications about the contaminants present on the polymer surface, however its poor sensitivity makes it unsuitable to analyse weakly contaminated samples. Thanks to its higher sensitivity and spatial resolution, ToF-SIMS analyses provide a deeper insight, which include also morphological information.
Images produced by ToF-SIMS 
confirm the presence of metal impurities implanted on the surface of several polymeric samples.

\begin{figure}[ht]
    \subfloat[\label{fig:subfigHSSdoppiaA}]{
    \includegraphics [angle=0, width=.46\columnwidth]{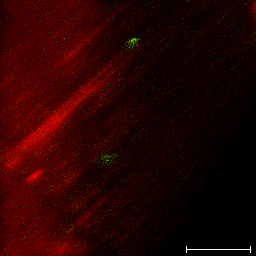}}
    \hfill
    \subfloat[\label{fig:subfigHSSdoppiaB}]{
    \includegraphics [angle=0, width=.46\columnwidth]{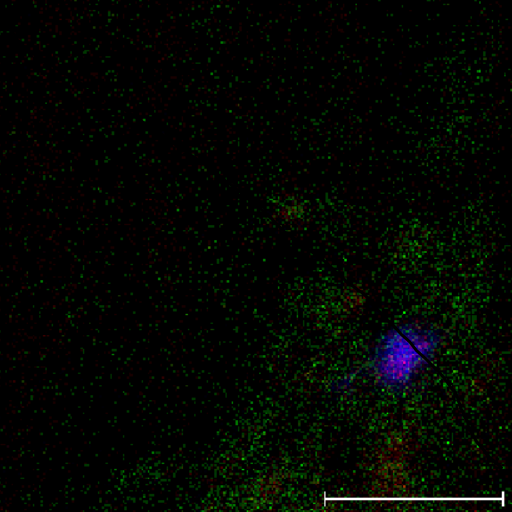}}
    \caption{ToF-SIMS images of submillimetric portions of PEEK surface machined with conventional tools (high-speed-steel, HSS). The data bars measure 100 $\mu$m. The image at left shows the total ion concentration (in red) in overlay together with the iron image (in green). Two well defined iron spots appear, denoting the presence of localized particles about 15-20~$\mu$m in size. In other instances different iron distribution are recorded. In the case shown in the image at right, the iron contaminant is spread over a wider area (about 200 $\mu$m in diameter), around a 20~$\mu$m spot containing a high concentration of Al (in blue) and Si (in red). }
\label{fig:HSSdoppia}
\end{figure}

Repeated ToF-SIMS maps performed on similar samples confirm the presence of iron (and other metals) contamination, often appearing in micrometric fragments (see Fig\ref{fig:subfigHSSdoppiaA}). They also  show that the contamination may occur with a scarce reproducibility. As an example, in Fig.\ref{fig:subfigHSSdoppiaB}, a spread Fe contamination interests an area about 200$\mu$m in diameter, around a localized micrometric spot with a high concentration of Al and Si.

In conclusion, ToF-SIMS analyses confirm the surface nature of the ferromagnetic contamination, also in accordance with information provided by the PEEK  producer \cite{privateVictrex18}: no ferromagnetic substances are used in the PEEK  production process, so that their presence in traces can only be attributed to residual impurities.

However,  as discussed in the following (see Secs.\ref{sec:titanium} and \ref{sec:cleaning}), there is also an evidence that some volume contamination level is also present. Compared to  surface contamination, such bulk term is much weaker, to the point of making it tolerable for our applications. 

Several approaches have been attempted to prevent or to remove the surface contamination, and their effectiveness have been evaluated on the basis of the above mentioned techniques. They are described extensively in the Secs.\ref{sec:titanium} and \ref{sec:cleaning}.

\section{Preventing surface contamination: \correction{non-ferromagnetic} tools}

\label{sec:titanium}

The surface contamination  most probably occurs at the machining stage. To prove that, titanium tools have been crafted and used at the lathing stage to avoid direct transfer of contaminants from the blade to the PEEK surface.

In this way  two additional evidences emerged proving that  surface contamination occurs during the manufacturing process. The first one is indirect: polymeric samples machined with titanium tools instead of ordinary HSS tools give much weaker DMFV. The second evidence is direct and is provided by ToF-SIMS maps, where no micrometric iron spots appear, while a barely detectable, homogeneously spread iron contamination is sometimes visible.

\begin{figure}[ht]
    \centering
    \subfloat[\label{fig:noFeinTiSmpl}]{
    \includegraphics [angle=0, width=.48\columnwidth]{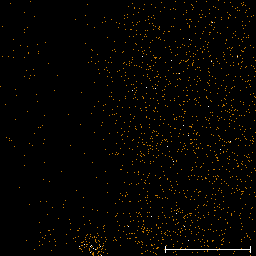}}
    \hspace{.07\columnwidth}
    \subfloat[\label{fig:TiSpoiled}]{
    \includegraphics [angle=0, width=.28 \columnwidth]{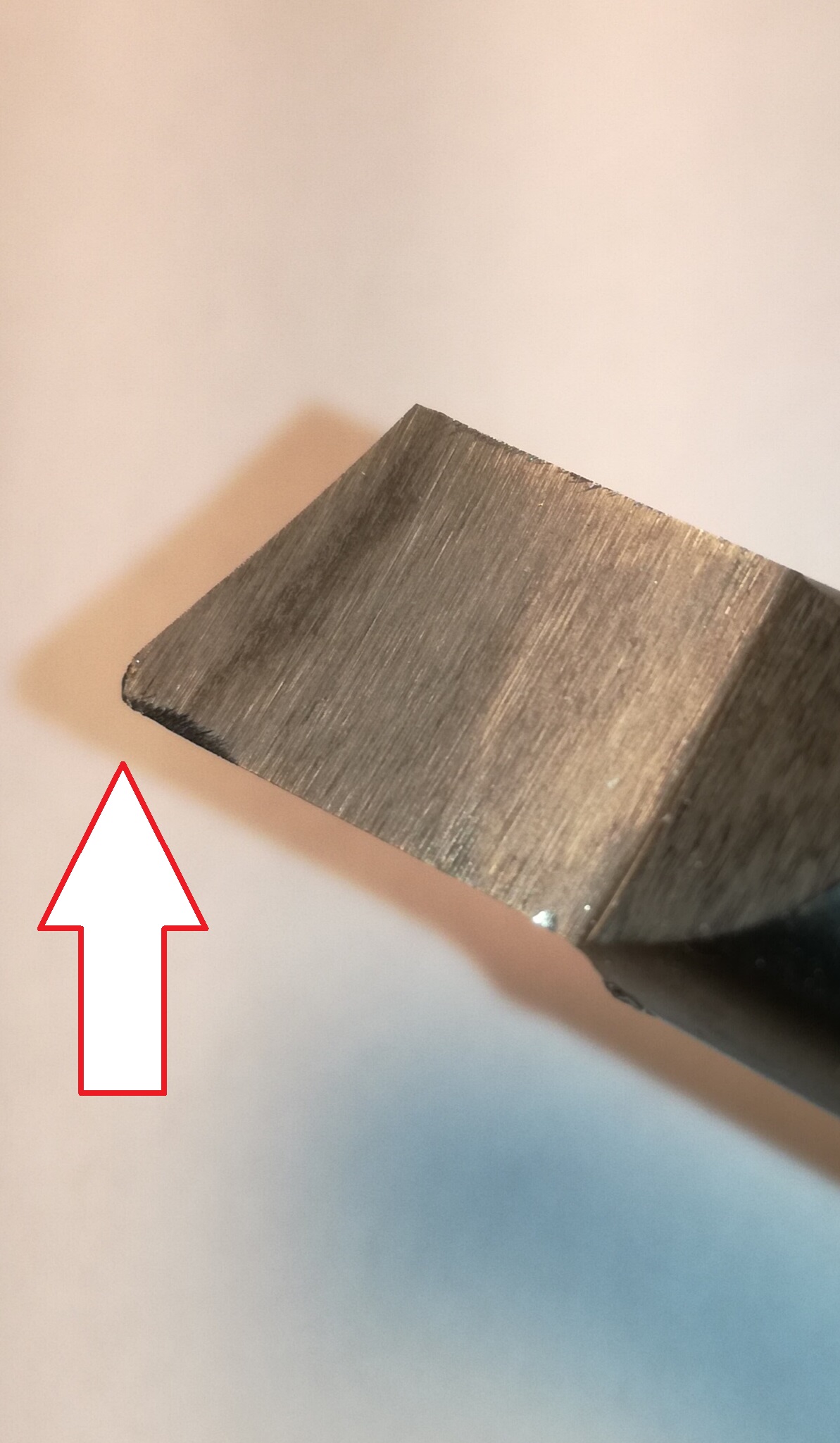}}
    \caption{(a) Iron map from ToF-SIMS of a submillimetric portion of PEEK surface machined with a Ti tool. The data bar measures 100 $\mu$m. The spread iron contamination appears at ppb level, and might be due just to instrumentation noise, in analogy with Al and Ti maps.  (b) The Ti tool blade spoiled after having machined two PEEK cylinders.}
\label{fig:Titanium}
\end{figure}

Titanium-machined samples result in a much weaker static field perturbation (DMFV $\approx \pm2 \text{ and } \pm4\,$nT for cartridges and cylinders, respectively).
Noticeably, this residual signal is proportional to the polymer volume and not to its surface.
Correspondingly and consistently --in the case of cartridges-- a narrower spread of proton precession frequencies is recorded.

These low DMFV levels approach the sensitivity limit (the precision is $ \approx\pm200$~pT). This is the lowest DMFV value obtained in term of  magnetic contamination and it is the same value measued with another method that removes the cartridge contamination discussed in the next section. This fact, combined with the observed reversed ratio between DMFV from low contaminated cartridge and cylinder, is a reliable indication that the residual level of ferromagnetism comes from the material bulk.

Concerning the ToF-SIMS measurements, no iron microparticles appear in Ti-worked samples. A very weak concentration (at ppb level, close to the instrumental sensitivity) of sub-pixel size iron signal is re\-cord\-ed, as shown in Fig.\ref{fig:noFeinTiSmpl}. 
Due to the poorer hardness and resilience, the Ti tools sharpness is spoiled in  a much shorter time with respect to the steel (see Fig.\ref{fig:TiSpoiled}) \correction{\cite{Victrex_tools}}, suggesting that much more particles are ripped out. Surprisingly, polymeric samples worked  with Ti tools result uncontaminated also in terms of Ti particles, when analyzed chemically (both Al and Ti maps do not present spots in Ti machined samples). A possible explanation is that the morphology of the ripped Ti fragments does not facilitate their penetration in the PEEK surface. 

Summarizing, the use of non-magnetic machining tools is a promising and effective approach to prevent the ferromagnetic contamination of cartridges. 

The titanium tools do not allow to make screwed caps, making necessary the use of glue to seal the cartridges. From a practical point of view, the excellent thermal and mechanical properties of PEEK facilitate the re-use of glued-cap cartridges. The cap can be removed heating up the aqueous content in a commercial microwave oven, so to produce an over pressure and to explode the container.

Glass lathing tools were also tested, with results as good as those obtained with titanium in terms of contamination, but with the drawback of an even worse resilience.
\correction{Another attempt was made with a tungsten carbide tool as specifically recommended for PEEK machining \cite{Victrex_tools}. This tool did not suffer from the wearing shown by the Ti one, but  the magnetic contamination resulted similar to the HSS case. }

\section{Removing surface contamination: cleaning methods}
\label{sec:cleaning}
An alternative procedure to reduce the surface contamination is the use of mechanical or chemical methods to remove residues from HSS-machining.

\subsection{Abrasion cleaning}
\label{subsec:abrasione}
The use of appropriate (magnetically tested) sand-paper to polish  the cylinder surface after its HSS-machining helps in  reducing the ferromagnetic response. 
The procedure is barely reproducible, and (crucial point) sand paper polishing is not practicable in the inner surface of cartridges, for which, consequently, alternative methods have to be developed in view of application. However, the effect is evident and this suggests that an intensive mechanical abrasion might constitute an valid method to produce clean containers.

In this perspective, a mechanical (abrasion) cleaning approach based on a tumbling technique was attempted and tested. Both cylinders and cartridges with threaded caps were treated  in a tumbling machine with corundum (Al$_2$O$_3$) sand, which is commercially available and is commonly used, e.g., for jewelry polishing. Different tumbling durations were applied, ranging from ten hours to several days.  

The  tumbling technique produces evident results, however the contaminant reduction does not achieve a satisfactory level. In the case of cylinders,
20 hours tumbling reduce the signal by 10\%  (DMFV $\approx \pm18\,$nT), and further 20\% (DMFV $\approx \pm15\,$nT) is obtained after additional 20 hours. A long lasting (250 h) tumbling, resulted in a 40\% reduction DMFV (from $\approx \pm16\,$nT to $\approx \pm10\,$nT):
an appreciable but yet improvable level, despite an evident polishing of the surface, that after such long treatment appeared glossy to the naked eye and perfectly smooth to the touch.

\begin{figure}[ht]
    \subfloat[]{\includegraphics [angle=0, width=  .46\columnwidth] {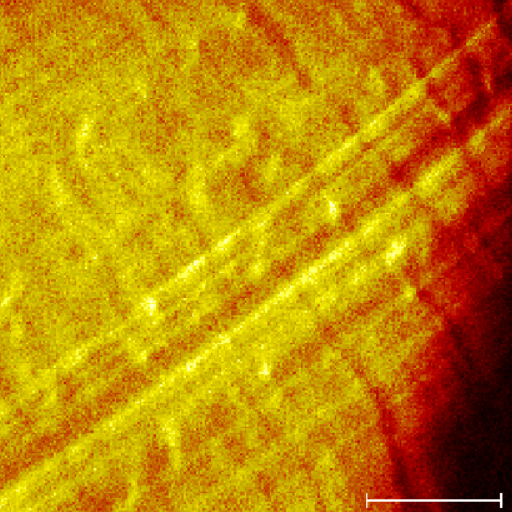}}
    \hspace{.07\columnwidth}
    \subfloat[]{\includegraphics [angle=0, width=  .46\columnwidth] {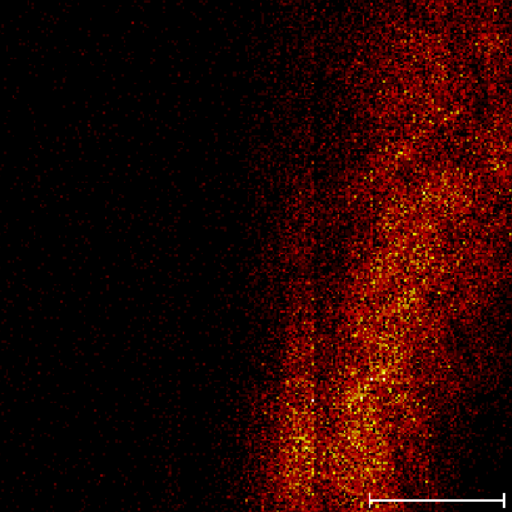}}
    \hfill
    \subfloat[]{\includegraphics [angle=0, width=  .46\columnwidth] {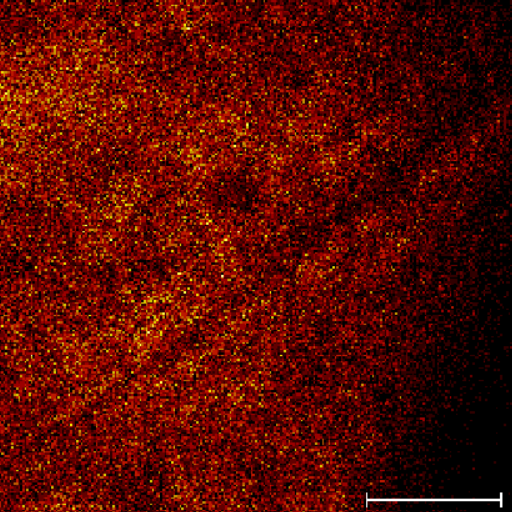}}
    \hspace{.07\columnwidth}
    \subfloat[]{\includegraphics [angle=0, width=  .46\columnwidth] {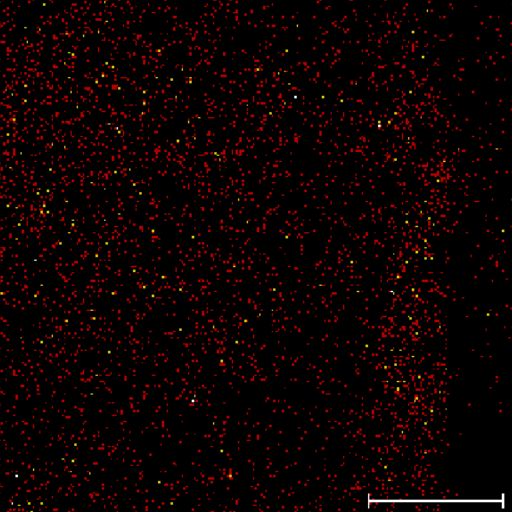}}
    \caption{
	ToF-SIMS Chemical maps of cartridge cleaned by abrasion. The image (a) shows the total ion image, the  (b) is the Al map, the (c) is the Si map, and the (d) is the Fe map. The data bars measure 100 $\mu$m. Iron contamination persists, but no microparticles appear. Distributed Si and Al contamination is evident all over the surface. Noticeably, the near-edge regions are more heavily contaminated by Al,  likely due to corundum residues, implanted where the tumbling is more effective. 
	\label{fig:B}}
\end{figure}

These results are consistent with the ToF-SIMS analysis performed on a tumbled sample. As shown in Fig.\ref{fig:B}, there is an evidence of residual iron  contamination but no large size (micrometric) particles are visible anymore, an additional, persistent Si and Al contamination is pointed out. The latter is more evident where the mechanical action of the tumbling is stronger (in the proximity of the edges). This indicates that some corundum particles penetrate the surface as to produce Al contamination.

In conclusion, both magnetometric and ToF-SIMS results indicate that the tumbling approach is not a promising method. It definitely provides some degree of cleaning, but it requires a long-lasting treatment and the final DMFV maintains unsatisfactorily high levels.

\subsection{Chemical cleaning}
\label{subsec:acidi}
Chemical cleaning based on acids has been attempted as well. Cylinders have been treated for different durations with several kinds of acids, namely 
\begin{itemize}
\item H$_2$SO$_4$ 8M;
\item H$_2$SO$_4$ 8M / H$_2$O$_2$ 10\%;
\item HNO$_3$ 0.5 M (PEEK does not allow exposure to higher concentrations of nitric acid);
\item oxalic acid (H$_2$C$_2$O$_4$), 
\end{itemize}

These cylinders have then been analysed both magnetometrically and by XRFS or ToF-SIMS. As summarized in Table \ref{tab:tabelladue}, the magnetometric measurements showed evident yet unsatisfactory decontamination levels, with the only exception of the oxalic acid.   Oxalic acid exhibited the best capacity to remove Fe from the polymeric surface, consistently with several previous  studies available in the literature \cite{panias_hydromet_96,lee_metall_07,salmimies_ccm_11}.

\begin{table}[ht]
\centering
\begin{tabular}{ | m{28mm} | m{10mm} | m{18mm} | } 
\hline
 \bf treatment & \bf time & \bf DMFV  \\ 
             &     &  \bf reduction  \\ 
\hline
 H$_2$SO$_4$ 8M & 3h & 
$50 \%$ \\ 
\hline
H$_2$SO$_4$ / H$_2$O$_2$ 10\% & 3h &
35\% \\ 
\hline
HNO$_3$ 0.5 M  & 3h  & 
8\% \\ 
\hline
H$_2$C$_2$O$_4$ 0.3M  & 24h & 
73\% \\ 
\hline
\hline
\end{tabular}
\caption{Signals from different PEEK cylinders before and after treatments of different duration. The only acid treatment leading to satisfying results is with oxalic acid (H$_2$C$_2$O$_4$).}
\label{tab:tabelladue}
\end{table}

\begin{figure}[ht]
    \subfloat[]{\includegraphics [angle=0, width=  .31\columnwidth] {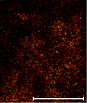}}
    \hspace{.02\columnwidth}
    \subfloat[]{\includegraphics [angle=0, width=  .31\columnwidth] {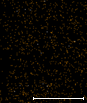}}
    \hspace{.02\columnwidth}
    \subfloat[]{\includegraphics [angle=0, width=  .31\columnwidth] {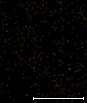}}
    \caption{
	Iron maps of PEEK disks lathed by ordinary HSS tools to fit the ToF-SIMS holder; the data bars measure 100 $\mu$m. The maps are recorded without any purification (a), after 1 hour sonication in 0.3M oxalic acid (b), and after an overnight treatment (c). In the last case a reduction of the iron contamination down to the detection threshold is achieved.
	\label{fig:sm3}}
\end{figure}

An overnight treatment in oxalic acid associated with sonication removes the contamination below the ToF-SIMS detection level and reduces the DMFV down to $\pm$4~nT. A set of ToF-SIMS maps corresponding to oxalic acid treatments is shown in Fig.\ref{fig:sm3}.

An example of measurement sets performed prior and after oxalic treatment has been shown in Fig.\ref{fig:hists}, where the second line (b, d, f) of histograms denote the improvement achieved in terms of the three indicators: decay time spread and mean value, DMFV and precession frequency spread. Quite similar results are obtained with Ti machined cartridges.
\\
The fact that the two techniques produce similar residual DMFV reinforces the hypothesis formulated in Sec.\ref{sec:titanium} that this persistent level is most likely due to a volume contamination.

In this respect, for comparison, we have measured the DMFV in cylinders made of other dielectric materials, observing smaller -however, still well detectable- values. In particular, an Acrylonitrile-butadiene-styrene (ABS) cylinder produced by a 3D extrusion printer and an Acetale cylinder lathed with amagnetic tool show about a $\pm$800~pT and $\pm$1~nT DMFV, respectively.

\section{Conclusion}
\label{sec:conclusioni}
The problem of residual permanent magnetization of polymeric sample containers used in a remote-detection ULF-NMR experiment has been studied by means of magnetometric measurement and chemical analyses of the polymer surface.

Some degree of magnetic remanence has been pointed out, and has been attributed  to polymer surface contamination occurring during the container production, and (at a much lower extent, as to have negligible consequences in our applications) to bulk contamination of the material.
The latter evidence emerges, at comparable levels, also with other polymeric materials. Consistently,  presence of contaminants in extruded polymers has been recently observed \cite{franco_pol_19}.

Despite the demonstrated existence of ferromagnetic behaviours in polymers \cite{mahmood_chem_18}, in this case the residual bulk ferromagnetism is attributable to diluted contaminants (at ppm concentration), dispersed in the volume during the machining stages (extrusion) of the  production of the PEEK rods. It can be matter of micro-particles dispersed in the volume, which are not detected by the ToF-SIMS analyses of clean samples, due to the extremely low probability of appearing in the small surface portions analysed or to nano-particles in  extremely dilute concentration, which appear in ToF-SIMS close to the detectable level.

The cartridge magnetization causes spurious fields both outside and inside the cartridges. These fields are evaluated magnetometrically both directly as a static dipolar field and \textit{via} their effects on the contained NMR sample. Complementary information is inferred from XRFS and ToF-SIMS analyses.

Several approaches have been proposed to counteract the ferromagnetic contamination and their effectiveness has been tested, by means of DC-magnetometry, ULF-NMR, XRFS and ToF-SIMS. 

Good results have been obtained by using non-mag\-net\-ic tools at the machining stage. This method poses some limitations in the cartridge construction. It reduces the ferromagnetic contamination to a level that is yet detectable, but eventually meets the requirements for accurate ULF-NMR spectroscopy.

Several post-production cleaning procedures have been tested, based on mechanical or chemical approaches.

Positive results have been obtained with sand-paper polishing (not applicable to thread and inner surfaces of the cartridges) and (to an unsatisfactory level) using tumbling techniques. Similarly, an appreciable contaminant reduction is observed with chemical treatments based on several strongly reactive acids, but the effect results insufficient even after long-lasting procedures. 

Excellent results are instead obtained with overnight oxalic acid treatment associated with sonication. In this case, the residual average remanence is the same as that achieved with non-magnetic machining, and there is an evidence that such level is attributable to ferromagnetic contaminants dispersed in the polymer volume at a sub ppm concentration.

\section{Acknowledgments}
The authors are pleased to thank Dr. Tommaso Lisini from SirsLab at DIISM, for providing the 3D printed samples, and Dr. Sophie Versavaud - Victrex Technical Support for the kind and effective assistance.

\section{Compliance with ethical standards}
\textbf{Conflict of Interest}
The authors declare that no conflict of interest exists, concerning the findings and the results presented in this paper.

\bibliographystyle{elsarticle-num} 
\bibliography{chemicontamrefs}

\begin{thebibliography}{10}
\expandafter\ifx\csname url\endcsname\relax
  \def\url#1{\texttt{#1}}\fi
\expandafter\ifx\csname urlprefix\endcsname\relax\def\urlprefix{URL }\fi
\expandafter\ifx\csname href\endcsname\relax
  \def\href#1#2{#2} \def\path#1{#1}\fi

\bibitem{wapler_jmr_14}
M.~C. Wapler, J.~Leupold, I.~Dragonu, D.~von Elverfeld, M.~Zaitsev,
  U.~Wallrabe, Magnetic properties of materials for {MR} engineering,
  micro-{MR} and beyond, Journal of Magnetic Resonance 242 (2014) 233 -- 242.

\bibitem{doty_cmr_98}
F.~D. Doty, G.~Entzminger, Y.~A. Yang, Magnetism in high-resolution {NMR} probe
  design. {I}: General methods, Concepts in Magnetic Resonance 10~(3) (1998)
  133--156.

\bibitem{bennett_jap_96}
L.~H. Bennett, P.~S. Wang, M.~J. Donahue, Artifacts in magnetic resonance
  imaging from metals, Journal of Applied Physics 79~(8) (1996) 4712--4714.

\bibitem{moessle_jmr_06}
M.~M{\"{o}}{\ss}le, S.-I. Han, W.~R. Myers, S.-K. Lee, N.~Kelso, M.~Hatridge,
  A.~Pines, J.~Clarke, {SQUID}-detected micro-{T}esla {MRI} in the presence of
  metal, Journal of Magnetic Resonance 179~(1) (2006) 146 -- 151.

\bibitem{zampetoulas_jmr_17}
V.~Zampetoulas, D.~J. Lurie, L.~M. Broche, Correction of environmental magnetic
  fields for the acquisition of nuclear magnetic relaxation dispersion profiles
  below earth's field, Journal of Magnetic Resonance 282 (2017) 38 -- 46.

\bibitem{biancalana_arnmrs_13}
G.~{Bevilacqua}, V.~{Biancalana}, Y.~{Dancheva}, L.~{Moi}, Chapter three -
  {Optical} atomic magnetometry for ultra-low-field {NMR} detection, in: G.~A.
  Webb (Ed.), Annual Reports on NMR Spectroscopy, Vol.~78 of Annual Reports on
  NMR Spectroscopy, Academic Press, 2013, Ch.~3, pp. 103 -- 148.

\bibitem{tayler_rsi_17}
M.~C.~D. Tayler, T.~Theis, T.~F. Sjolander, J.~W. Blanchard, A.~Kentner,
  S.~Pustelny, A.~Pines, D.~Budker, Invited review article: {Instrumentation}
  for nuclear magnetic resonance in zero and ultralow magnetic field, Review of
  Scientific Instruments 88~(9) (2017) 091101.

\bibitem{barskiy_nat_19}
D.~A. Barskiy, M.~C. Tayler, I.~Marco-Rius, J.~Kurhanewicz, D.~B. Vigneron,
  S.~Cikrikci, A.~Aydogdu, M.~Reh, A.~N. Pravdivtsev, J.-B. H{\"o}vener,
  et~al., Zero-field nuclear magnetic resonance of chemically exchanging
  systems, Nature communications 10~(1) (2019) 1--9.

\bibitem{biancalana_rsi_19}
G.~Bevilacqua, V.~Biancalana, Y.~Dancheva, L.~Stiaccini, A.~Vigilante, Spurious
  ferromagnetic remanence detected by hybrid magnetometer, Review of Scientific
  Instruments 90~(4) (2019) 046106.

\bibitem{tayler_apl_19}
M.~C.~D. Tayler, J.~Ward-Williams, L.~F. Gladden, Ultralow-field nuclear
  magnetic resonance of liquids confined in ferromagnetic and paramagnetic
  materials, Applied Physics Letters 115~(7) (2019) 072409.

\bibitem{biancalana_DH_jpcl_17}
G.~Bevilacqua, V.~Biancalana, Y.~Dancheva, A.~Vigilante, A.~Donati, C.~Rossi,
  Simultaneous detection of {H} and {D} {NMR} signals in a micro-{T}esla field,
  The Journal of Physical Chemistry Letters 8 (2017) 6176--6179.

\bibitem{biancalana_IDEA_prappl_19}
G.~Bevilacqua, V.~Biancalana, Y.~Dancheva, A.~Vigilante, Restoring narrow
  linewidth to a gradient-broadened magnetic resonance by inhomogeneous
  dressing, Phys. Rev. Applied 11 (2019) 024049.

\bibitem{biancalana_apl_19}
G.~Bevilacqua, V.~Biancalana, Y.~Dancheva, A.~Vigilante, Sub-millimetric
  ultra-low-field {MRI} detected in situ by a dressed atomic magnetometer,
  Applied Physics Letters 115~(17) (2019) 174102.

\bibitem{sepioni_epl_12}
M.~Sepioni, R.~R. Nair, I.-L. Tsai, A.~K. Geim, I.~V. Grigorieva, Revealing
  common artifacts due to ferromagnetic inclusions in highly oriented pyrolytic
  graphite, {EPL} (Europhysics Letters) 97~(4) (2012) 47001.

\bibitem{wang_jap_14}
Y.~Wang, X.~Chen, L.~Li, A.~Shalimov, W.~Tong, S.~Prucnal, F.~Munnik, Z.~Yang,
  W.~Skorupa, M.~Helm, S.~Zhou, Structural and magnetic properties of
  irradiated {S}i{C}, Journal of Applied Physics 115~(17) (2014).

\bibitem{kuroda_nat_14}
K.~Shinji, N.~Nozomi, T.~Koki, M.~Masanori, B.~Yoshio, O.~Krzysztof, T.~Dietl,
  Origin and control of high-temperature ferromagnetism in semiconductors,
  Nature Materials 6 (2007) 440--446.

\bibitem{zhao_prb_08}
G.~Zhao, P.~Beeli, Observation of an ultrahigh-temperature ferromagnetic-like
  transition in iron-contaminated multiwalled carbon nanotube mats, Phys. Rev.
  B 77 (2008) 245433.

\bibitem{muluaka_ijomeh_03}
M.~Muluaka, Assessment of lung particle accumulation in factory workers by
  magnetic field measurement, International Journal Occupation Medicine and
  Environmental Health (2003) 209--213.

\bibitem{pereira_jpd_11}
L.~M.~C. Pereira, J.~P. Ara{\'{u}}jo, M.~J.~V. Bael, K.~Temst, A.~Vantomme,
  Practical limits for detection of ferromagnetism using highly sensitive
  magnetometry techniques, Journal of Physics D: Applied Physics 44~(21) (2011)
  215001.

\bibitem{halbach_nim_80}
K.~Halbach, Design of permanent multipole magnets with oriented rare earth
  cobalt material, Nuclear Instruments and Methods 169~(1) (1980) 1 -- 10.

\bibitem{biancalana_apb_16}
G.~Bevilacqua, V.~Biancalana, P.~Chessa, Y.~Dancheva, Multichannel optical
  atomic magnetometer operating in unshielded environment, Applied Physics B
  122~(4) (2016) 103.

\bibitem{biancalana_rsi_14}
V.~Biancalana, Y.~Dancheva, L.~Stiaccini, Note: {A} fast pneumatic
  sample-shuttle with attenuated shocks, Review of Scientific Instruments
  85~(3) (2014).

\bibitem{biancalana_rsi_17}
V.~Biancalana, G.~Bevilacqua, P.~Chessa, Y.~Dancheva, R.~Cecchi, L.~Stiaccini,
  A low noise modular current source for stable magnetic field control, Review
  of Scientific Instruments 88~(3) (2017) 035107.

\bibitem{biancalana_rsi_10}
J.~{Belfi}, G.~{Bevilacqua}, V.~{Biancalana}, R.~{Cecchi}, Y.~{Dancheva},
  L.~{Moi}, {Stray magnetic field compensation with a scalar atomic
  magnetometer}, Review of Scientific Instruments 81~(6) (2010) 065103.

\bibitem{biancalana_FPGAstab_prappl_19}
G.~Bevilacqua, V.~Biancalana, Y.~Dancheva, A.~Vigilante, Self-adaptive loop for
  external-disturbance reduction in a differential measurement setup, Phys.
  Rev. Applied 11 (2019) 014029.

\bibitem{bellandbloom_prl_61}
W.~E. Bell, A.~L. Bloom, Optically driven spin precession, Phys. Rev. Lett. 6
  (1961) 280--281.

\bibitem{biancalana_zulfJcoupling_jmr_16}
G.~Bevilacqua, V.~Biancalana, A.~Ben Amar~Baranga, Y.~Dancheva, C.~Rossi,
  Micro-{T}esla {NMR} {J}-coupling spectroscopy with an unshielded atomic
  magnetometer, Journal of Magnetic Resonance 263 (2016) 65--70.

\bibitem{budker_nat_07}
D.~Budker, M.~Romalis, Optical magnetometry, Nature Physics (2007) 227--234.

\bibitem{biancalana_jmr_09}
G.~{Bevilacqua}, V.~{Biancalana}, Y.~{Dancheva}, L.~{Moi}, {All-optical
  magnetometry for {NMR} detection in a micro-{T}esla field and unshielded
  environment}, Journal of Magnetic Resonance 201 (2009) 222--229.

\bibitem{lifeng_mpemr_12}
L.~Zhang, J.~Gao, L.~N.~W. Damoah, D.~G. Robertson, Removal of iron from
  aluminum: A review, Mineral Processing and Extractive Metallurgy Review
  33~(2) (2012) 99--157.

\bibitem{kurtz_biomat_07}
S.~M. Kurtz, J.~N. Devine, {{PEEK} biomaterials in trauma, orthopedic, and
  spinal implants}, Biomaterials 28 (2007) 4845--4869.

\bibitem{bertocco_ieee_94}
M.~Bertocco, C.~Offelli, D.~Petri, Analysis of damped sinusoidal signals via a
  frequency-domain interpolation algorithm, IEEE Transactions on
  Instrumentation and Measurement 43~(2) (1994) 245--250.

\bibitem{yoshida_jpe_82}
I.~{Yoshida}, T.~{Sugai}, S.~{Tani}, M.~{Motegi}, K.~{Minamida}, H.~{Hayakawa},
  {Automation of internal friction measurement apparatus of inverted torsion
  pendulum type}, Journal of Physics E Scientific Instruments 14 (1981)
  1201--1206.

\bibitem{leone_cons_19}
G.~Leone, A.~D. Vita, M.~Consumi, G.~Tamasi, C.~Bonechi, A.~Donati, C.~Rossi,
  A.~Magnani, Comparison of original and modern mortars at the herculaneum
  archaeological site, Conservation and Management of Archaeological Sites
  21~(2) (2019) 92--112.

\bibitem{leone_ass_12}
G.~Leone, M.~Consumi, S.~Lamponi, A.~Magnani, Combination of static time of
  flight secondary ion mass spectrometry and infrared reflection-adsorption
  spectroscopy for the characterisation of a four steps built-up carbohydrate
  array, Applied Surface Science 258~(17) (2012) 6302 -- 6315.

\bibitem{caciagli_jmmm_18}
A.~Caciagli, R.~J. Baars, A.~P. Philipse, B.~W. Kuipers, Exact expression for
  the magnetic field of a finite cylinder with arbitrary uniform magnetization,
  Journal of Magnetism and Magnetic Materials 456 (2018) 423 -- 432.

\bibitem{privateVictrex18}
V.~E. GmbH, {Technical Service - Victrex Technical Support}, Private
  communication (2018).

\bibitem{Victrex_tools}
{Victrex},
  \href{https://www.victrex.com/~/media/literature/en/victrex_finishing-brochure.pdf}{{Victrex
  Finishing Brochure}} (2016).
\newline\urlprefix\url{https://www.victrex.com/~/media/literature/en/victrex_finishing-brochure.pdf}

\bibitem{panias_hydromet_96}
D.~Panias, M.~Taxiarchou, I.~Paspaliaris, A.~Kontopoulos, Mechanisms of
  dissolution of iron oxides in aqueous oxalic acid solutions, Hydrometallurgy
  42~(2) (1996) 257 -- 265.

\bibitem{lee_metall_07}
S.~O. Lee, T.~Tran, B.~H. Jung, S.~J. Kim, M.~J. Kim, Dissolution of iron oxide
  using oxalic acid, Hydrometallurgy 87~(3) (2007) 91 -- 99.

\bibitem{salmimies_ccm_11}
R.~Salmimies, M.~Mannila, J.~Kallas, A.~H{\"a}kkinen, Acidic dissolution of
  magnetite: Experimental study on the effects of acid concentration and
  temperature, Clays and Clay Minerals 59~(2) (2011) 136--146.

\bibitem{franco_pol_19}
M.~F. Franco, R.~Gadioli, M.~A. De~Paoli, {Presence of iron in polymers
  extruded with corrosive contaminants or abrasive fillers}, {Pol\'imeros} 29
  (2019).

\bibitem{mahmood_chem_18}
J.~Mahmood, J.~Park, D.~Shin, H.-J. Choi, J.-M. Seo, J.-W. Yoo, J.-B. Baek,
  Organic ferromagnetism: Trapping spins in the glassy state of an organic
  network structure, Chem 4~(10) (2018) 2357--2369.

\end{thebibliography}

\end{document}